# Pulse delay via tunable white light cavities using fiber optic resonators

**H.N. Yum[1,3,*], Y.J. Jang[1], M.E. Kim[2], X. Liu[1], M.S. Shahriar[1,2]**

[1]*Department of Electrical Engineering and Computer Science, Northwestern University, Evanston, IL 60208*
[2]*Department of Physics and Astronomy, Northwestern University, Evanston, IL 60208*
[3]*Department of Electrical Engineering, Texas A&M University, College Station, TX 77843*
[*]*Corresponding author:* h-yum@northwestern.edu

**Abstract**

Previously, we proposed a data buffering system that makes use of a pair of white light cavities[1]. For application to telecommunication systems, it would be convenient to realize such a device using fiber optic resonators. In this paper, we present the design of such a system, where the white light cavity effect is produced by using stimulated Brillouin scattering. The system consists of a pair of fiber optic white light cavities placed in series. As in the original proposal, the delay time can be controlled independently of the bandwidth of the data pulses. Furthermore, we show how the bandwidth of the system can be made as large as several times the Brillouin frequency shift. We also show that the net delay achievable in such a buffer can be significantly larger than what can be achieved using a conventional recirculating loop buffer.

Slow-light in optical fibers has been of interest due to its applicability to current optical devices for fiber optic communication such as optical buffers, optical delay lines and fast memory access [2,3,4,5,6]. However, the amount of delay achieved is typically too small to be of interest for most applications. Recently, we have shown that this limitation can be overcome by using fast-light, in a manner that is rather counter-intuitive[1]. Briefly, this approach makes use of so-called white light cavities (WLC's). A WLC is a cavity containing a fast-light medium, tuned so that negative dispersion causes the wavelength to become independent of frequency over a certain spectral range. As such, it resonates over a broader spectral range compared to an empty cavity of equal length and finesse, without a reduction in the cavity build-up factor[7]. The buffer system is composed of two WLCs as well as an intervening zone of dispersion-free propagation. When the fast-light medium is deactivated, the WLC acts as a narrow-band cavity, which reflects a high-bandwidth pulse stream. However, when the fast-light medium is activated, the data stream passes through the WLC. Using these properties, the data stream can be trapped between the two WLCs for a duration which is limited only by the residual transmission through the cavity in the narrow-band mode, and the length of the intervening zone. As shown in ref. 1, such a buffer can slow down a data pulse for a duration several thousand times longer than the pulse, with virtually no distortion. However, for many reasons, a buffer of this type based on free space components is likely to be impractical, especially for telecommunication. In this paper, we show how to realize such a buffer using WLC's base of fiber resonators, with an optical fiber forming the intervening path.

In some ways, the buffer proposed here is similar in configuration to the feedback buffer employing a recirculation loop[8,9,10,11]. However, the key difference is that once the data is in the loop, it is almost completely isolated. During each circulation through the loop, the attenuation is due to a vanishingly small coupling to the WLC, and the residual transmission loss inherent to the fiber. As such, there is no need for an amplifier in the loop. A single stage amplification upon release from the buffer is sufficient to restore the signal level to the input value. Elimination of an intra-loop amplifier entails absence of noise due to amplified spontaneous emission, so that for a given level of signal to noise ratio (SNR), a much larger number of loop circulations can be allowed. Furthermore, absence of intra-loop amplification reduces the energy cost of the buffer.

Before proceeding with the analysis of a fiber-based fast-light buffer, it is instructive to consider first the basic building block: a fiber resonator coupled to an optical fiber, as illustrated in Fig.1(a). The model presented here is based on the general configuration of a ring resonator coupled to a waveguide[12,13,14,15,16]. We assume that the 2X2 coupler is internally lossless. Thus, the complex amplitudes $a_i$ and $b_i$ are related simply by the intensity coupling coefficient $k$:

$$\begin{bmatrix} b_1 \\ b_2 \end{bmatrix} = \begin{bmatrix} \sqrt{1-k} & j\sqrt{k} \\ j\sqrt{k} & \sqrt{1-k} \end{bmatrix} \begin{bmatrix} a_1 \\ a_2 \end{bmatrix} \quad (1)$$

Next, we express the transmission within the ring resonator as $a_2 = \alpha e^{j\theta} b_2$ in terms of the transmission factor $\alpha$ through the fiber loop and the round trip phase shift $\theta$, which can be expressed as $-\omega n(\omega) L/c$ where $L$ is the circumference of the ring resonator and $n(\omega)$ is the refractive index of the fiber. Using Eq. (1), we then get:

$$\frac{b_1}{a_1} = \frac{\sqrt{1-k} - \alpha e^{j\theta}}{1 - \alpha\sqrt{1-k} e^{j\theta}} \quad (2)$$

$$\frac{a_2}{a_1} = \frac{j\alpha e^{j\theta} \sqrt{k}}{1 - \alpha e^{j\theta} \sqrt{1-k}} \quad (3)$$

Fig.1(b) displays a fiber resonator coupled to a single mode fiber through the second coupler. It will be a building block for our proposed fiber-based data buffering system. The fiber-coupled resonator can be treated effectively as an uncoupled resonator with additional loss. Therefore, the fields $a_i$, $b_i$ (i=1,2) are related by the same matrix as presented in Eq.(1), provided that $k$ is replaced by $k_1$ to represent the coupling coefficient for the first coupler, and $\alpha$ is replaced by $\alpha\sqrt{1-k_2}$, where $k_2$ is the coupling coefficient of the second coupler. We thus get:

$$\frac{b_1}{a_1} = \frac{\sqrt{1-k_1} - \alpha\sqrt{1-k_2} e^{j\theta}}{1 - \alpha\sqrt{1-k_1}\sqrt{1-k_2} e^{j\theta}} \quad (4)$$

$$\frac{a_2}{a_1} = \frac{j\alpha\sqrt{k_1}\sqrt{1-k_2} e^{j\theta}}{1 - \alpha\sqrt{1-k_1}\sqrt{1-k_2} e^{j\theta}} \quad (5)$$

In addition, the following relations hold:

$$a_2 = \alpha\sqrt{1-k_2} e^{j\theta} b_2 \quad (6a)$$

$$a_r = j\sqrt{\alpha}\sqrt{k_2} e^{\frac{\theta}{2}j} b_2 \quad (6b)$$

Combining Eqs. 5 and 6, we get:

$$\frac{a_r}{a_1} = \frac{-\sqrt{\alpha}\sqrt{k_1}\sqrt{k_2} e^{j\frac{\theta}{2}}}{1 - \alpha\sqrt{1-k_1}\sqrt{1-k_2} e^{j\theta}} \quad (7)$$

In order to take into account dispersion in the fiber loop (induced by SBS, for example), we express $n(\omega)$ in terms of a Taylor expansion about $\omega_0$: $n(\omega) = n_0 + (\omega - \omega_0)n_1 + (\omega - \omega_0)^3 n_3$, $n_1 = dn/d\omega|_{\omega=\omega_0}$, $n_3 = (1/6) d^3n/d\omega^3|_{\omega=\omega_0}$ where $n_0$ is the mean index of the fiber.

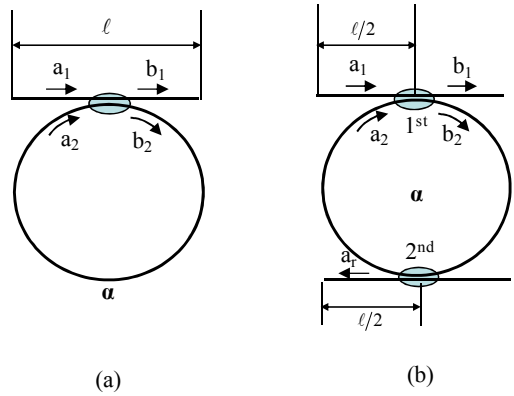

Fig. 1 Schematics of (a) fiber ring resonator, (b) ring resonator coupled to a single mode fiber

Fig. 2 displays the numerical simulations for $|a_r/a_1|^2$ of a fiber-coupled resonator in the absence of dispersion (blue line, $n_1 = 0$, $n_3 = 0$) and with negative dispersion (red line, $n_1 < 0$, $n_3 \neq 0$). For simplicity, we assume unit input intensity $|a_1| = 1$ and no internal loss ($\alpha = 1$). We choose $n_1$ to fulfill the ideal White Light Cavity (WLC) condition[7]. For the configuration presented in Fig.1(b), the length of the dispersive medium is assumed to be equal to that of the ring resonator. In that case, the ideal WLC condition requires $n_g = 0$, where $n_g$ is the group index of the dispersive fiber[7]. Next, $n_3$ is adjusted so that the WLC linewidth becomes finite. We have chosen $k_1 = k_2 = 0.01$, $\ell = 1$, $n_0 = 1.45$ and $L=10.6897$ where $L$ and $\ell$ are in meter. We have used $\omega_0 = 2\pi \times 1.9355 \times 10^{14}$, corresponding to 1550nm in telecommunication bandwidth. As can be seen from Fig. 2, the linewidth of WLC is expanded, compared to the ordinary ring resonator associated with $n(\omega) = n_0$. It should be noted that this broadening occurs without a reduction in the cavity build-up factor[7]. If the WLC linewidth is broad enough for the pulse spectrum to be under the resonant spectral region of WLC, then the input signal will transmit without loss or distortion.

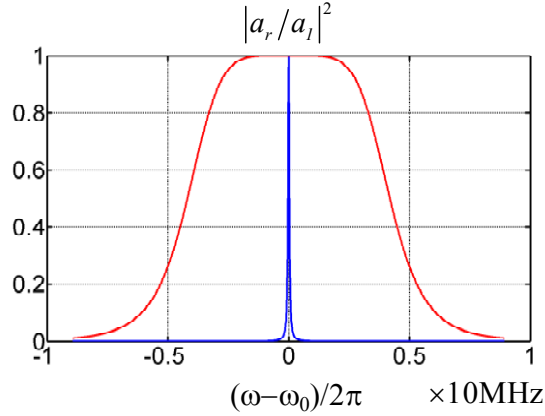

Fig. 2 $|a_r/a_I|^2$ for the fiber-coupled resonator presented in Fig. 1(b). The blue trace is for the case where the resonator is assumed to be non-dispersive, while the red trace is for the case where it is anomalously dispersive ($n_1 = -1.192 \times 10^{-15} / rad$, $n_3 = 1.223 \times 10^{-32} / rad^3$)

Next, we consider the propagation of a pulse through such a fiber-coupled resonator. Eq. 7 represents the transfer function between the input and the output. The transfer function is denoted as $H_0$ for the resonator without dispersion [ $n(\omega) = n_0$ ], and as $H_{WLC}$ under the white light cavity condition [ $n(\omega) = n_0 + (\omega - \omega_0)n_1 + (\omega - \omega_0)^3 n_3$ ]. To find the group velocity associated with the system, it is important to express the group index in terms of $\angle H_{0/WLC}$, the phase shift induced during propagation through the resonator. The phase contribution resulting from the propagation through the whole system (fiber plus cavity) can be expressed as $(\omega n_{eff} \ell)/c = (n_0 \omega \ell)/c - \angle H_{0/WLC}$, where we define $n_{eff}$ for the effective refractive index provided by the resonator. By the definition of the group index, obviously $n_{g(resonator)} = n_{eff} + \omega(dn_{eff}/d\omega)$ where $n_{g(resonator)}$ is the group index of the whole system (and not the group index of the medium). Thus, $n_{g(resonator)}$ is given by[14]

$$n_{g(resonator)} = n_0 - \frac{c}{\ell}\frac{d\angle H_{0/WLC}}{d\omega} \quad (8)$$

Note that the pulse distortion would be characterized[1] by $\Delta T \approx -\left(d^2 \angle H_{0/WLC}/d\omega^2\right)\Delta\omega$.

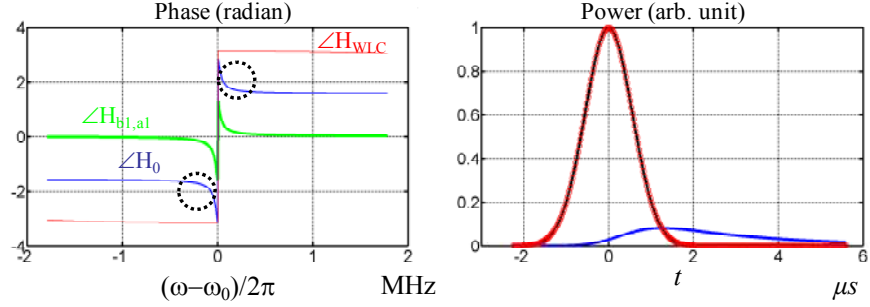

Fig. 3 (a) Phases associated with the transfer functions of the resonator displayed in Fig. 1(b). $\angle H_{b1,a1} = arg(b_1/a_1)$, $\angle H_0 = arg(a_r/a_1)$ in the absence of WLC effect, and $\angle H_{WLC} = arg(a_r/a_1)$ for WLC, with $n_1 = -1.192 \times 10^{-15} / rad$, $n_3 = 1.223 \times 10^{-32} / rad^3$ (b) Reference pulse after propagating in a fiber of length $\ell$ (black) and the outputs associated with $H_0$ (blue) and $H_{WLC}$ (red). The output in the presence of the WLC effect is essentially overlapped with the reference.

It is instructive to compare $\angle H_{0/WLC}$ to the phase of $b_1/a_1$ in Eq. (4), denoted as $\angle H_{a1,b1}$ in Fig. 3(a). Fig. 3(b) graphically shows the output pulses resulting from propagation through the system, in the presence of the cavities associated with $H_0$ (blue) and $H_{WLC}$ (red circles) as well as the pulse after propagating a distance $\ell$ through a fiber only without dispersion (black). For illustration, we used the cavity parameters in Fig. 2 and chose the input pulse to be of the form $S_{in}(t) = exp(-t^2/t_0^2)exp[j(\omega_0 + \xi)t]$. Here, $t_0$ is chosen so that $\Delta\nu_{pulse} = 10\Delta\nu_{cavity}$, where $\Delta\nu_{pulse} = 1/t_0$, and $\Delta\nu_{cavity}$ is the full-width-half-maximum (FWHM) of the ordinary resonator. Fourier transform of $S_{in}(t)$ leads us to $\tilde{S}_{in}(\omega) = t_0/\sqrt{2}\, exp[-\{(\omega - \omega_0 - \xi)t_0\}^2/4]$. Applying the convolution theorem, we obtain the amplitude of the output pulse as: $S_{out}(t) = 1/\sqrt{2} \int_{-\infty}^{\infty} S(\omega)exp(-jk_0\ell)H_{0/WLC}(\omega)exp(j\omega t)d\omega$, where $k_0 = n_0\omega\ell/c$. We simply set $H_{0/WLC} = 1$ when the field propagates in the fiber only.

Note that there is a discontinuity accompanied by a phase leap at $\omega = \omega_0$ as illustrated in Fig.3(a). Of course, such a discontinuity disappears when the sources of all losses as well as the finite bandwidth of a real signal are taken into account. However, under the assumptions used here, this result can be explained as follows, in analogy with the critically coupled microresonator presented in ref. 13, for example. Specifically, a critically coupled resonator shows a $\pi$-phase leap on resonance. For the resonator considered here, we have used $\alpha = 1$, $k_1 = k_2 = 0.01$ so that $\alpha\sqrt{1-k_2} = \sqrt{1-k_1}$. This means that the transmission factor between $b_2$ and $a_2$ ($\alpha\sqrt{1-k_2}$) matches the transmission coefficient of the first coupler ($\sqrt{1-k_1}$), corresponding to critical coupling. Thus, $\angle H_{b1,a1}$ shows the $\pi$-phase leap at resonance. The second coupler, which is identical to the first one, is also critically coupled. Thus, the $\pi$-phase leap occurs twice, resulting in a discontinuity of $2\pi$ for $\angle H_{0/WLC}$.

We explain the output pulses illustrated in Fig. 3(b) with the aid of Fig. 3(a). According to Eq.(8), the negative slope of $\angle H_0$ suggests $n_{g(resonator)} > n_0$ inside the dotted circle. By setting $\xi = 0$, we have chosen the input pulse to have the carrier frequency equal to $\omega_0$. As such, the pulse lies within the slow light zone. Since $\Delta\nu_{pulse} = 10\Delta\nu_{cavity}$, most of the pulse spectrum is under the spectral region of $|H_0| = 0$. As a consequence, the output associated with $H_0$ is delayed and attenuated, as can be seen in Fig. 3(b). For the case of WLC, we consider $\xi = 1.5/t_0$ so as to ensure that the pulse spectrum is mostly outside the region where $\angle H_{WLC}$ leaps by $2\pi$. Thus, over the spectrum of the pulse, we have $|H_{WLC}| \simeq 1$, $d\angle H_{WLC}/d\omega \simeq 0$, and $d^2 \angle H_{WLC}/d\omega^2 \simeq 0$, so that $n_{g(WLC)} \simeq n_0$, according to Eq. (8). As a result, the output of WLC is not advanced compared to the reference pulse propagating the distance of $\ell$ through a bare fiber; rather, these outputs are virtually superimposed on each other, as illustrated in Fig. 3b. This behavior can also be understood physically noting that $n_g = 0$ for the fiber inside the resonator under the ideal WLC condition. Thus the pulse propagates in the resonator with the speed of $v_g \gg c$, thereby spending very little time inside. Of course, under realistic condition, such a propagation does not violate special relativity or causality[17]. In ref.18, we describe in detail the exact behavior of a pulse inside a cavity loaded with an anomalously dispersive medium, under a range of conditions, including $n_g = 0$.

In analogy with the previously proposed Fabry-Perot(FP) buffer system[1], we now present the design of a fiber-based data buffer, as shown in Fig. 4. We assume that a bi-frequency Brillouin pump creates a negative dispersion in a ring resonator to produce the WLC effect, in a manner analogous to the previous WLC demonstration where a bi-frequency Raman pump was used to produce dual Raman gain peaks[7], yielding a negative dispersion between the peaks. Here, each Brillouin pump produces a Brillouin gain peak for the counter-propagating probe. In the spectral region between these two gain peaks, the probe experiences negative dispersion, which can be tuned to reach the WLC condition. The WLC on the left (LWLC) is connected to the WLC on the right (RWLC) through fiber spools to construct a closed loop where a pulse would be trapped.

For a data pulse and the WLCs, we use the same parameters as considered in Fig.3. The operating scheme to delay the pulse without distortion is similar to that presented in ref. 1. When the pulse enters from left, we turn on the bi-frequency Brillouin pumps to activate the WLC effect in LWLC. Thus, the pulse transmits through the resonator with no distortion, as shown in Fig. 3(b). Once the pulse has left LWLC, we turn off the WLC effect. Now, the pulse is loaded and circulates inside the trapping loop.

To consider power loss during the circulation, it is important to note that the carrier frequency of the pulse is shifted by $\xi = 1.5/t_0$ from the resonant frequency, $\omega_0$, of the bare cavity (i.e., without the WLC effect). If it is not shifted, the pulse spectrum would include the transmission window of the bare cavities. In that case, the spectral component within this window would leak out through the bare cavities. With the carrier frequency shifted, LWLC as well as RWLC acts as a simple coupler, with an intensity coupling coefficient of $k_1 = k_2 = 0.01$. On each bounce, the pulse is reflected, with a small transmission loss due to this finite coupling coefficient.

Once we are ready to extract the pulse from the trapping loop, we activate the WLC effect in RWLC. On arriving at RWLC, the pulse passes through it with neither attenuation nor distortion.

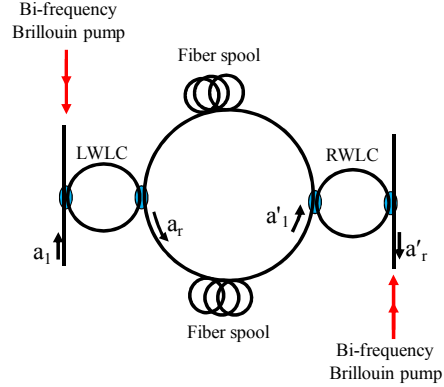

Fig. 4 Schematic illustration of the fiber-based data buffer system, employing bi-frequency Brillouin pumps. LWLC: Left White Light Cavity; RWLC: Right White Light Cavity.

To find a full transfer function to describe the data buffer system, we consider the amplitude transfer characteristics $H_{lr} = a_r/a_l$ ($H_{l'r'} = a'_r/a'_l$) where $H_{lr}$ ($H_{l'r'}$) denotes the transfer function of LWLC(RWLC). After N multiple round trips inside the trapping loop, $a_r$ is related to $a'_l$ by

$$H_{rl'} = \left(\sqrt{1-k_1}\sqrt{1-k_2}\right)^N e^{-jk_0\frac{2N+1}{2}L_2} 10^{-\frac{\alpha}{20}} \quad (9)$$

where $L_2$ is the length of the trapping loop and $k_0 = \omega n_0/c$. Here, α represents the total attenuation due to the propagation through the loop. A conventional single mode fiber for 1550nm exhibits an attenuation loss of ~0.2dB/$km$ so that $\alpha = 0.2\times(N+1/2)L_2$. The time elapsed in the loop represents the system delay: $\tau_d = [n_0(N+1/2)L_2]/c$. Fig.5 illustrates the reference pulse propagating a distance $\ell$ though a fiber, as well as the output from the data buffer for N=50. The reference pulse can be written as $S_{ref}(t) = 1/\sqrt{2}\int_{-\infty}^{\infty} exp(-jk_0\ell)S(\omega)exp(j\omega t)d\omega$. Using $H_{lr}$, $H_{l'r'}$ together with Eq.(9), the output of the data buffer can be written as:

$$S_{out}(t) = 1/\sqrt{2}\int_{-\infty}^{\infty} exp(-jk_0\ell)H_{l'r'}(\omega)H_{rl'}(\omega)H_{lr}(\omega)S(\omega)exp(j\omega t)d\omega \quad (10)$$

Figs. 5(a) and 5(b) show that for $L_2 = 500$ the output pulses is delayed by $1.22\times10^{-4}$ seconds compared to the reference. Fig. 5(b) indicates that the pulse delay is accompanied by no serious distortion, but an attenuation of about 9.3 dB (from 1 to 0.115). The attenuation per pass is 0.186 dB, which is the sum of attenuation due to transmission (0.1dB), plus a loss of 1% (0.043 dB) at each of the two couplers.

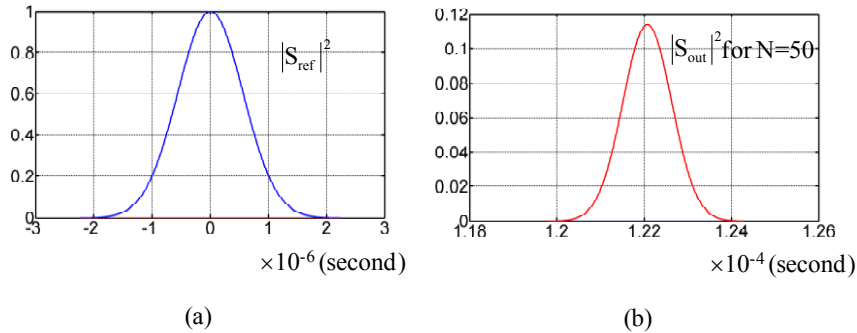

Fig. 5 Output pulses from the fiber-based buffering system: (a) reference pulse, (b) pulse after 50 round trips inside the trapping loop.

The attenuation suffered in the storage loop can be compensated by using an optical amplifier. Such an amplifier could be added to the storage loop. However, since the loss per pass in the loop is very small, and an amplifier would have an insertion loss much higher than the single pass attenuation, a better approach is to use a separate loop for the amplification. This is illustrated schematically in Fig.6(a). Prior to the buffering process, both WLCs are inactive. The pulse stream to be stored is inserted into the trapping loop by activating $WLC_1$, and deactivating it after the stream is fully loaded. We assume the perimeter of the trapping loop to be 0.5 *km* (same as the loop considered in figure 5). After 50 passes (with an attenuation of 9.3 dB), $WLC_2$ is activated, and the pulse stream enters the amplifying loop. The amplifier in this loop is gated to provide a net amplification (amplifier gain minus the insertion loss of the amplifier) of 9.3 dB, restoring the original amplitude. We neglect the attenuation in the amplifying loop, which can be much smaller. Once the stream re-enters the trapping loop, $WLC_2$ is deactivated. This process is repeated *M* times after another 50 passes through the trapping loop.

The number of times the amplification is applied, *M*, is limited by the fact that the signal to noise ratio (SNR) is degraded due to noise added during each pass through the amplification process[9]. The actual reduction in SNR during each pass would depend on the type of amplification employed. The maximum allowable net reduction in SNR would depend on the SNR in the input pulse stream, and the fidelity requirement of the system. As an example, we consider a case where *M* is limited to 100. The net delay achievable is then ~12.2 msec, and the delay-bandwidth product for the input pulse used in Fig 5 would be $10^4$. Obviously, if much shorter pulses are used (which would require a higher bandwidth WLC), the delay bandwidth product (DBP) can be correspondingly larger. For example, for a pulse width of 0.122 nsec (requiring a WLC linewidth of ~ 30GHz), DBP would be $10^8$.

It is instructive to compare such a system with a conventional recirculating buffer[8,9,10,11]. A typical implementation of such a buffer is illustrated in Fig.6(b). Here, during each pass, there is a loss of 7 dB due to the two couplers and the isolator. The net gain (amplifier gain minus the insertion loss of the amplifier) provided by the amplifier in each pass is thus 7 dB. If all other parameters are comparable to the buffer shown in Fig.6(a), then the maximum number of amplification for approximately the same reduction in SNR would be about 133 (=9.3×100/7). The net delay achievable would be 0.33 msec. Thus, all else being equal, the delay time achievable for the buffer proposed here would achieve a delay of nearly 37 times larger than what can be achieved using a conventional recirculating buffer. This is attributable solely to the fact that the conventional buffer has a large loss (7dB) per pass, while for the WLC buffer, the inherent loss per pass is much smaller (0.186 dB). The factor by which the WLC delay is larger is essentially a ratio of these two numbers. The relative advantage thus would become better as the loop perimeter becomes smaller, and/or couplers with higher efficiency are employed.

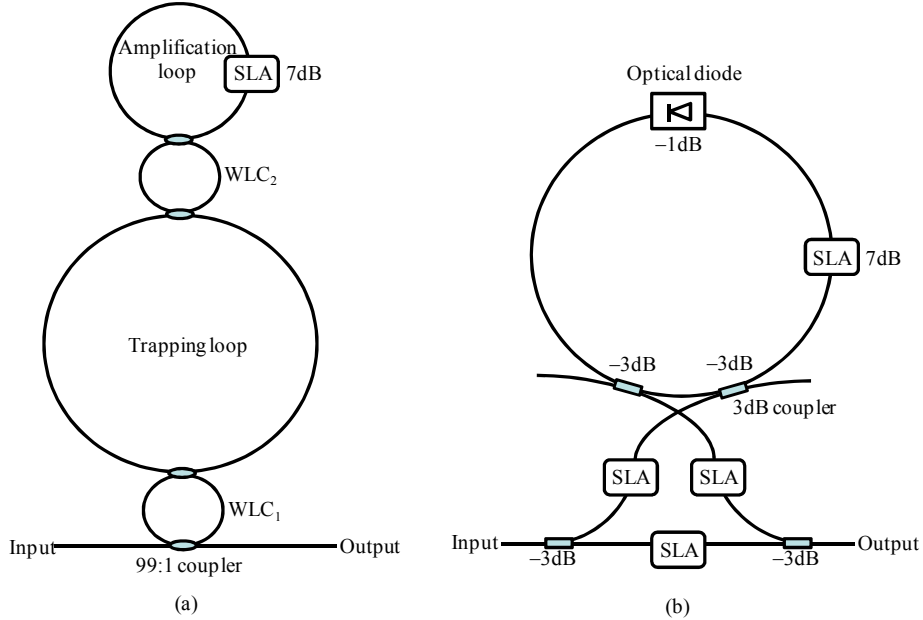

Fig. 6. Schematics illustration of (a) a WLC-based data buffer augmented by amplification, (b) a conventional recirculating data buffer[9]; SLA: Semiconductor Laser Amplifier; WLC$_{1,2}$: White Light Cavity$_{1,2}$. In Fig. 6(b), the optical diode acts as an isolator.

Next, we discuss a technique for producing a high bandwidth for the data buffer system. From the numerical simulations presented above, we find that in general $\Delta v_{WLC}/2$ corresponds to the system bandwidth, where $\Delta v_{WLC}$ is the WLC bandwidth. This is because the pulse spectrum is placed within either one of these two regions, in order to avoid transmission loss at the bare cavity resonance frequency in the trapping loop. The value of $\Delta v_{WLC}$ depends on two important parameters: $\delta$, the gain separation, and $\Delta v_B$, the linewidth of the gain profile. As mentioned above, a negative dispersion is created between the two gain profiles. Of course, the slope at the center of the two gains become smaller with decreasing $\Delta v_B$ for a particular $\delta$ or increasing $\delta$ for a particular $\Delta v_B$. However, the WLC condition requires a slope that yields a vanishing group index. Thus, for a given value of $\Delta v_B$, the value of $\delta$ needed for the WLC effect is fixed. This value of $\delta$ increases with increasing $\Delta v_B$. In ref. 7, we showed that $\Delta v_{WLC} \sim \delta^{2/3}$. Thus, in order to enhance $\Delta v_{WLC}$, it is necessary to increase $\Delta v_B$.

Broadening of the Brillouin gain profile can be achieved, for example, by superposing a Gaussian white noise on the dc current of a laser diode[19,20,21,22] used as a Brillouin pump. In particular, in ref. 22, two Brillouin pumps with equal power, each broadened by white noise, were separated by $2v_B$ (where $v_B$ is the Brillouin frequency shift) to produce a single gain with a bandwidth of $2v_B$. Here, we present a scheme to expand the spectral range of a negative dispersion in a similar manner, as illustrated in Fig. 7. We consider first two groups of Brillouin pumps. Each group consists of a pair of pumps with equal power, denoted as pumps $A_1$ & $A_2$ ($B_1$ & $B_2$). As shown in Fig. 7(a), pumps $A_1$ and $A_2$ ($B_1$ and $B_2$) are separated by $2v_B$, and pumps $A_1$ and $B_1$ ($A_2$ and $B_2$) by $\delta$. Fig. 7(b) indicates that the spectra of all pumps are broadened so that each pump produces a broadband Brillouin gain with $\Delta v_{gain} \sim v_B$.

Note that each pump with frequency $v_p$ generates an absorption profile at $v_p + v_B$ with an amplitude equal to that of the gain profile at $v_p - v_B$. As such, the loss spectra induced by pump $A_1$ (pump $B_1$) is compensated by the gain of pump $A_2$ (pump $B_2$), since equal-power pumps are used. As displayed in Fig. 7(c), for a single group, it is possible to increase $\Delta v_B$ until the tail of the gain profile produced by pump $A_1$ (pump $B_1$) meets that of the absorption of pump $A_2$ (pump $B_2$). Accordingly, we get $\Delta v_B \simeq 2v_B$.

In what follows, we assume $v' \simeq 2\Delta v_B$ where $v'$ is the width measured along the bottom of the gain. For the two groups of pumps, it is important to consider the overlap of the net loss profile of $A_2$ with the gain profile of $B_1$. Note that, as illustrated in Fig. 7(d), these two profiles meet in the encircled area if $\Delta v_B \simeq 2v_B$. Such an overlap distorts the net gain profile. In order to avoid it, the parameters under consideration should satisfy the condition that $\delta + v'/2 \leq 2v_B$.

In principle[22], N pumps can create a single gain with a maximum bandwidth equal to $Nv_B$. In that case, $2Nv_B$ is the spectral distance between the gain peak of pump $A_1$ and the absorption dip of pump $A_N$. To assure that the tail of the gain profile of pump $B_1$ does not encounter that of the absorption profile of pump $A_N$, the condition is $\delta + v'/2 \leq 2(N-1)v_B$. Fig. 8 displays the gain profiles for a particular condition: N=3, $\Delta v_B = 2v_B$, $\delta = 2v_B$, and $v' = 4v_B$. Fig. 8(a) indicates that the gain of pump $A_{N+1}(B_{N+1})$ counters the loss due to pump $A_N(B_N)$ (N=1,2). In Fig.8(b), the gain profiles of pump $A_1$ and $B_1$ remain, and thus the net gain profile is a gain doublet with the separation of $\delta = 2v_B$. The Brillouin frequency of conventional optical fibers is 8~12GHz[23,24]. Thus, for example, with N=11, $v_B = 10$GHz, and $\Delta v_B = 10v_B$, it is possible to get $\delta = 100$GHz and $\Delta v_{WLC} = 21.5$GHz.

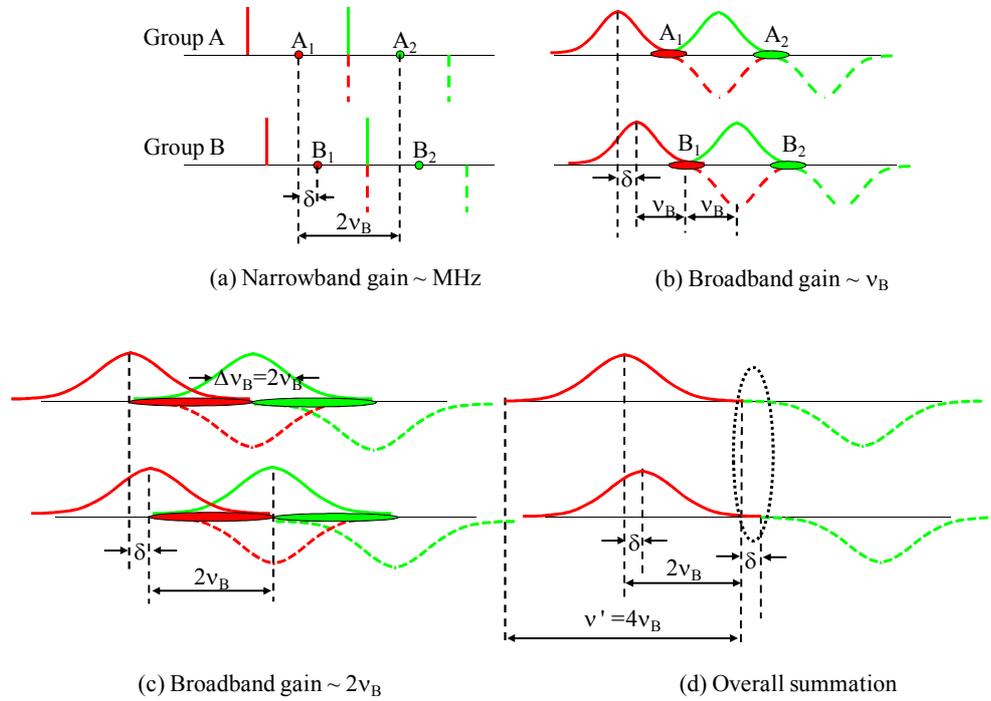

Fig. 7. Illustration of a scheme to increase $\delta$ and $\Delta\nu_B$ using broad-band Brillouin pumps. (a) In the absence of Gaussian white noise, the pumps produce the narrowband Brillouin gain (solid bar)/loss (dotted bar) spectrum. (b) The linewidth of the pumps is broadened by the white noise. The gain bandwidth is expanded to $\nu_B$, (c) and eventually reaches the maximum of $2\nu_B$. (d) In the encircled area, the net gain in group B is overlapped with the net loss in group A.

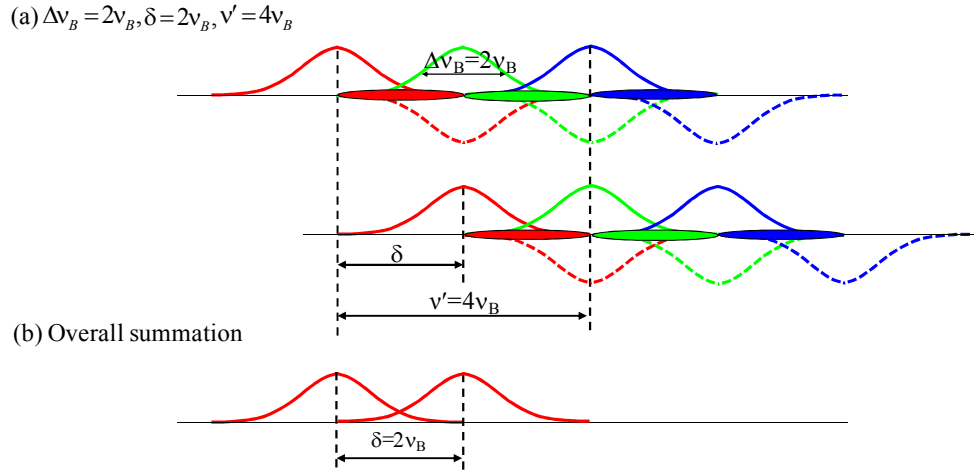

Fig. 8 (a) Each group consists of three broad-band pumps. (b) Net gain profile is a gain doublet with $\delta = 2\nu_B$.

To summarize, we present the design of fiber-based fast-light data buffer system, consisting of a pair of white light cavities placed in series. The white light cavity effect is produced by using stimulated Brillouin scattering. The delay time can be controlled independently of the bandwidth of the data pulses, thus circumventing the delay-bandwidth product constraint faced by conventional buffer systems. Furthermore, we show how the bandwidth of the system can be made as large several times the Brillouin frequency shift. We also show that the net delay achievable can be significantly larger than what can be achieved with a conventional recirculating loop buffer.

This work was supported by DARPA through the slow light program under grant FA9550-07-C-0030, by AFOSR under grants FA9550-06-1-0466 and FA9550-10-1-0228, and by the NSF IGERT program under grant number DGE-0801685.